\documentclass[twocolumn,showpacs,preprintnumbers,amsmath,amssymb]{revtex4}

\usepackage{graphicx}
\usepackage{dcolumn}
\usepackage{bm}

\begin{document}

\preprint{APS/123-QED}

\title{Precise half-life measurement of $^{110}$Sn and $^{109}$In isotopes}

\author{Gy.\,Gy\"urky}%
 \email{gyurky@atomki.hu}
\author{Z.\,Elekes}%
\author{Zs.\,F\"ul\"op}%
\author{G.G.\,Kiss}%
\author{E.\,Somorjai}%
\affiliation{%
Institute of Nuclear Research (ATOMKI), H-4001 Debrecen, POB.51., Hungary}%
\author{A.\,Palumbo}%
\author{M.\,Wiescher}
\affiliation{%
University of Notre Dame, Notre Dame, Indiana 46556, USA}

\date{\today}

\begin{abstract}
The half-lives of $^{110}$Sn and $^{109}$In isotopes have been
measured with high precision. The results are
T$_{1/2}$\,=\,4.173\,$\pm$\,0.023\,h for $^{110}$Sn and
T$_{1/2}$\,=\,4.167\,$\pm$\,0.018\,h for $^{109}$In. The precision
of the half-lives has been increased by a factor of 5 with respect
to the literature values which makes results of the recently
measured $^{106}$Cd($\alpha,\gamma$)$^{110}$Sn and
$^{106}$Cd($\alpha$,p)$^{109}$In cross sections more reliable.
\end{abstract}


\maketitle

The stellar nucleosynthesis process originating heavy, proton rich
isotopes - the so-called p-nuclei - is the astrophysical p-process
\cite{ar03}. The modeling of the p-process requires the knowledge
of the nuclear reaction rates associated with the p-process
reaction network. Due to lack of experimental data, these reaction
rates are generally calculated with Hauser-Feshbach statistical
model calculations \cite{NON-SMOKER,MOST} which use nuclear
physics inputs such as optical model potentials, nuclear level
densities, ground state properties, $\gamma$-ray strength
functions, etc. The ambiguities of these input parameters
introduce considerable uncertainties into the reaction rate
predictions and subsequently into the p-process model simulations.
The reliability of the calculations can be checked by comparison
with the experimental cross sections. This requires data in the
relevant stellar energy range, known as Gamow window. Of
particular relevance is the study of p-process reaction rates on
neutron deficient nuclei in the Z=50, N=50 mass range. These
reactions are associated with the production of the light
p-nuclei, $^{92,94}$Mo, $^{96,98}$Ru up to $^{113}$In. In these
cases p-process model predictions differ substantially from the
observed p-process abundances \cite{ar03}. While many alternative
scenarios have been offered to explain these discrepancies it is
clearly necessary to probe the reliability of the presently used
nuclear physics input in calculating the p-process reaction rates
in this mass range near the closed shell N=50 nuclei. Recently, a
number of experiments has been devoted to the study of
(p,$\gamma$) and ($\alpha,\gamma$) reaction cross sections at the
astrophysically relevant energies and the results are compared
with the model calculations (see e.g. \cite{gyu03}). Generally,
the models are able to reproduce the experimental results
reasonably well, however, deviations up to a factor of 2 have been
observed. Comparing the experimental data with the model
predictions could help find the best input parameter sets for the
statistical model.

The activation technique has developed as the most important tool
in (p,$\gamma$) and ($\alpha,\gamma$) p-process experiments
\cite{sa97,so98,bo98,ch99,gyu01,oz02,gyu03}. In this technique the
cross sections are determined by measuring the decay activity of
the reaction product. The resulting cross sections are directly
correlated with the half-life of the reaction product; deviations,
ambiguities, and uncertainties in available half-life data
translate directly into uncertainties of experimental cross
section results.

Recently, the $^{106}$Cd($\alpha,\gamma$)$^{110}$Sn and
$^{106}$Cd($\alpha$,p)$^{109}$In cross sections have been measured
in the energy range relevant for the astrophysical p-process
\cite{gyu05} and the results are compared with the model
calculations. In order to reduce the systematic error of the
measurement, the reaction has been measured independently in two
laboratories. All major error sources such as target thickness,
detector efficiency, charge collection, and counting statistics
are determined independently in the two measurements thereby
making the experimental results more reliable. The decay
parameters such as the half-life of the reaction products,
however, enter the analysis of both measurements. The error of the
half-life is thus reflected directly in the error of the derived
cross sections.

The half-lives of the reaction products found in the literature
are T$_{1/2,adopted}$($^{110}$Sn)\,=\,4.11\,$\pm$\,0.10\,h
\cite{NDS89} and
T$_{1/2,adopted}$($^{109}$In)\,=\,4.2\,$\pm$\,0.10\,h
\cite{NDS86}. These half-life values have relatively large errors
and are based on measurements carried out many decades ago. The
half-life of $^{110}$Sn is based on two experiments; one is only
available as unpublished thesis result \cite{me56} while only very
limited experimental detail is provided by the second work
\cite{ka73}. The situation is somewhat better in the case of $^{109}$In, where
the compilation is based on four works \cite{ma49,mc51,no62,sm68}. However, the quoted 
half-lives have large errors and there is no detailed discussion about the experimental 
setup and data analysis in those works.
An independent confirmation of these results is in particular
necessary to provide reliable reference data for the
$^{106}$Cd($\alpha,\gamma$)$^{110}$Sn and
$^{106}$Cd($\alpha$,p)$^{109}$In activation measurements. The aim
of the present work is to check the reliability of the adopted
half-life values and to reduce their errors.

In the present experiment the half-lives of $^{110}$Sn and
$^{109}$In isotopes have been measured simultaneously. The sources
were produced by bombarding a $^{106}$Cd target with a 12.4\,MeV
$\alpha$-beam at the cyclotron laboratory in ATOMKI, Deb\-re\-cen.
The target was prepared by evaporating highly enriched metallic
$^{106}$Cd (96.47\,\% enrichment) onto a 3\,$\mu$m thick Al foil.
The thickness of the target was roughly 400\,$\mu$g/cm$^2$. The
$\alpha$ beam intensity was about 500\,nA and the irradiation
lasted for 10 hours. The size of the beam spot and hence the size
of the source was roughly 8\,mm in diameter. The $^{110}$Sn
isotope was produced by the $^{106}$Cd($\alpha,\gamma$)$^{110}$Sn
reaction. There are two ways to produce the $^{109}$In isotope:
directly by the $^{106}$Cd($\alpha$,p)$^{109}$In reaction or since
$^{109}$Sn decays to $^{109}$In, by the
$^{106}$Cd($\alpha$,n)$^{109}$Sn reaction. In order not to have
additional feeding to $^{109}$In during the half-life measurement
(which might distort the result), the half-life of $^{109}$In can
be determined precisely only after $^{109}$Sn has decayed
completely (T$_{1/2}$\,=\,18.0\,$\pm$\,0.2\,m); therefore, the
gamma-counting has been started 6 hours after the end of the
irradiation.

\begin{figure}
\resizebox{\columnwidth}{!}{\rotatebox{270}{\includegraphics{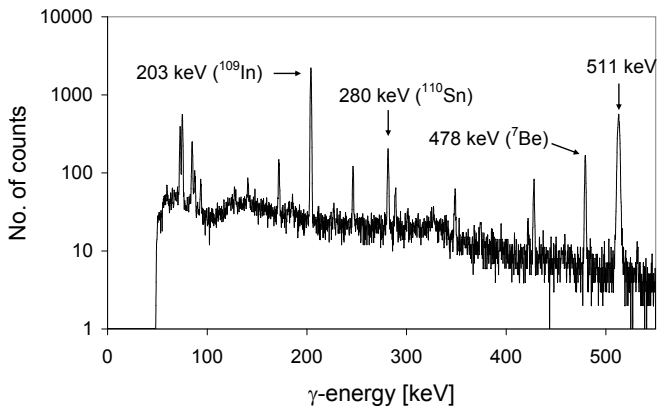}}}
\caption{\label{fig:spec}Gamma spectrum taken 10 hours after the
start of the counting in a counting interval of 15 minutes. The
lines used for the analysis (see text) together with the strong
511~keV annihilation line are indicated. The other visible peaks
are either from laboratory background or peaks of induced
radioactivity on impurities in the target.}
\end{figure}

The irradiated sample has been placed in front of a 40\% relative
efficiency HPGe detector in a holder fixed rigidly onto the end
cap of the detector. Directly at the back of the sample a $^7$Be
source has been put in the holder in order to be able to control
any possible change in the detection geometry or the efficiency of
the detector during the counting. The system was shielded by 5~cm
thick lead. The decay of the $^{110}$Sn and $^{109}$In isotopes
has been followed for 24 hours (roughly 6 half-lives) recording
the $\gamma$-spectra in every 15 minutes. Altogether 96 spectra
were collected. Fig.~\ref{fig:spec} shows a typical
$\gamma$-spectrum.

The electron capture decay of $^{110}$Sn is followed by the
emission of a single 280\,keV $\gamma$-radiation with 100\,\%
relative intensity. The detection of this line was used to deduce
the half-life of $^{110}$Sn. At the beginning of the counting the
intensity of this line was roughly 6000\,counts in 15 minutes
which went down to about 100 counts in 15 minutes by the end of
the measurement. The strongest $\gamma$-radiation following the
$\beta$-decay of $^{109}$In is the 203 keV line which has a
relative intensity of 74\,\%. This line was used for the analysis.
Its intensity reduced from about 67000 counts/15\,min to 1200
counts/15\,min in the course of the measurement.

The spectra were collected with 100~MHz Wilkinson type ADC with
8192 channel MCA having a built-in dead time correction. Owing to
the very low counting rate, the dead time was always very low
ranging from an initial value of $\approx$ 1\,\% declining to a
value of $\approx$~0.1\,\% by the end of the counting. Dead time
corrections have been made for the final decay time analysis. To
determine the effect of the dead time uncertainties, the decay
time was determined assuming an initial dead-time between 2\,\% and
0\,\%. This changes the final half-life value by only $\approx$~0.3\,\%; 
this uncertainty has been incorporated in the final error
of the derived half-life (for the not normalized value, see Table.
\ref{tab:result}).

\begin{table}
\caption{Obtained half-lives for the two measured isotopes}
\label{tab:result}
\begin{ruledtabular}
\begin{tabular}{lccc}
 & \multicolumn{3}{c}{Half-life [hours]} \\
 & without & normalized & weighted \\
 & normalization & to $^7$Be & average  \\
\hline \\
$^{110}$Sn & 4.179 $\pm$ 0.023 & 4.165 $\pm$ 0.035 & 4.173 $\pm$ 0.023 \\
$^{109}$In & 4.168 $\pm$ 0.018 & 4.166 $\pm$ 0.022 & 4.167 $\pm$ 0.018
\end{tabular}
\end{ruledtabular}
\end{table}

The half-lives of the investigated isotopes were determined also
by normalizing the intensities of the 203 and 280 keV peaks with
the number of counts in the 478\,keV peak coming from the decay of
the $^7$Be reference source. In this case the slight change of the
$^7$Be activity (T$_{1/2}$\,=\,53.22\,$\pm$\,0.06\,days,
\cite{NPA708}) during the one day $\gamma$-counting has been taken
into account. The counting rate of the 478\,keV $^7$Be peak was
about 900 counts in 15 minutes. The dead time correction affects
equally the number of counts in the $^7$Be and
$^{110}$Sn/$^{109}$In peaks; therefore, no dead time correction
needs to be applied in the case of the normalized half-lives.

\begin{figure}[h]
\resizebox{\columnwidth}{!}{\rotatebox{270}{\includegraphics{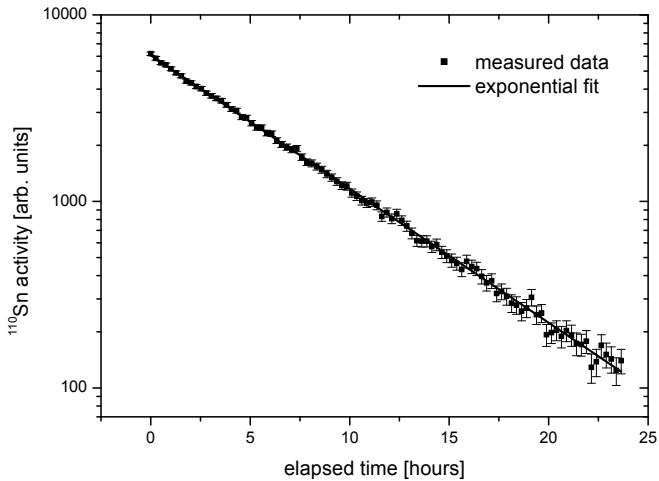}}}
\caption{\label{fig:decay}Decay of $^{110}$Sn measured for 24
hours. The points are the measured area of the 280 keV peak
without the normalization with $^7$Be. The solid line is the
exponential fit to the measurement.}
\end{figure}

In all cases the half-lives have been determined from the
parameters of the exponential fit to the measured data. As an
example, Fig.~\ref{fig:decay} shows the decay curve of the
$^{110}$Sn isotope without normalization.

Table~\ref{tab:result} shows the results for both isotopes
obtained without and with normalization to $^7$Be. The two methods
give the same result within the error bar for both isotopes
indicating that there is no long-term instability in the detection
system and the dead time correction is made properly. The weighted
averages are listed in the last column of the Table. Since the two
methods are statistically correlated, the error of the weighted
average was chosen to be the error of the non-normalized value.
Our final result for the half-life of $^{110}$Sn, {\bf T$_{{\bf
1/2}}$\,=\,4.173\,${\bf \pm}$\,0.023\,h}, is slightly longer than
the adopted value (T$_{1/2,\,adopted}$\,=\,4.11\,$\pm$\,0.10\,h)
in the data compilations. The error is reduced from 2.4 to
0.5\,\%. The obtained $^{109}$In  half-life of {\bf T$_{{\bf
1/2}}$\,=\,4.167\,${\bf \pm}$\,0.018\,h} is shorter than the
adopted value (T$_{1/2,\,adopted}$\,=\,4.2\,$\pm$\,0.10\,h) listed
in the compilations and the error is also strongly reduced.

Based on the new half-life values the uncertainty in the
$^{106}$Cd($\alpha,\gamma$)$^{110}$Sn and
$^{106}$Cd($\alpha$,p)$^{109}$In cross section measurements is
significantly reduced and the comparison with statistical model
calculations becomes more reliable.


This work was supported by OTKA (Grant Nos. T034259, T042733,
F043408 and D048283), by the NSF-Grant PHY01-40324 and through the
Joint Institute of Nuclear Astrophysics (www@JINAweb.org) NSF-PFC
grant PHY02-16783 . Gy.~Gy and Zs.~F acknowledge support from the
Bolyai grant.


\begin{thebibliography}{9}
\bibitem{ar03} M. Arnould and S. Goriely, Phys. Rep. {\bf 384} 1 (2003).
\bibitem{NON-SMOKER} T. Rauscher and F. K. Thielemann, At. Data Nucl. Data Tables {\bf 79}, 47 (2001).
\bibitem{MOST} S.\,Goriely, in Nuclei in the Cosmos V, Edition Fronti\`eres Paris, (1998), p. 314
\bibitem{gyu03} G. Gy\"urky, Zs. F\"ul\"op, E. Somorjai, M. Kokkoris, S. Galanopoulos, P. Demetriou, S. Harissopulos, T. Rauscher, and S. Goriely, Phys. Rev. C {\bf 68}, 055803 (2003).
\bibitem{sa97} T. Sauter and F. K\"appeler, Phys. Rev. C {\bf 55}, 3127 (1997).
\bibitem{so98} E. Somorjai, Zs. F\"ul\"op, \'A.Z. Kiss, C.E. Rolfs, H.P. Trautvetter, U. Greife, M. Junker, S. Goriely, M. Arnould, M. Rayet, T. Rauscher, and H. Oberhummer, Astron. Astrophys. {\bf 333} (1998) 1112.
\bibitem{bo98} J. Bork, H. Schatz, F. K\"appeler, and T. Rauscher, Phys. Rev. C {\bf 58}, 524 (1998).
\bibitem{ch99} F. R. Chloupek, A. StJ. Murphy, R. N. Boyd, A. L. Cole, J. G\"orres, R. T. Guray, G. Raimann, J. J. Zach, T. Rauscher, J. V. Schwarzenberg, P. Tischhauser, and M. C. Wiescher, Nucl. Phys. {\bf A652}, 391 (1999).
\bibitem{gyu01} G. Gy\"urky, E. Somorjai, Zs. F\"ul\"op, S. Harissopulos, P. Demetriou, and T. Rauscher, Phys. Rev. C {\bf 64}, 065803 (2001).
\bibitem{oz02} N. \"Ozkan, A. StJ. Murphy, R. N. Boyd, A. L. Cole, M. Famiano, R.T. G\"uray, M. Howard, L. Sahin, J.J. Zach, R. deHaan, J. G\"orres, M. C. Wiescher, M.S. Islam, and T. Rauscher, Nucl. Phys. {\bf A710}, 469 (2002).
\bibitem{gyu05} Gy.\,Gy\"urky Zs.\,F\"ul\"op, G.\,Kiss, Z.\,M\'at\'e, E.\,Somorjai, J.\,G\"orres, A.\,Palumbo,
M.\,Wiescher, D.\,Galaviz, A.\,Kretschmer, K.\,Sonnabend, A.\,Zilges, and T.\,Rauscher, Nucl. Phys. A in press (preliminary results)
\bibitem{NDS89} D.\,De\,Frenne and E. Jacobs, Nucl. Data Sheets {\bf 89}, 481 (2000).
\bibitem{NDS86} J.\,Blachot, Nucl. Data Sheets {\bf 86}, 505 (1999).
\bibitem{me56} W. Mead, Thesis, Univ.California (1956); UCRL-3488 (1956)
\bibitem{ka73} H.M.A.\,Karim, Radiochim. Acta {\bf 19}, 1 (1973).
\bibitem{ma49} E.C.\,Mallary and M.L.\,Pool, Phys. Rev. {\bf 76}, 1454 (1949).
\bibitem{mc51} C.L.\,McGinnis, Phys. Rev. {\bf 81}, 734 (1951).
\bibitem{no62} M.\,Nozawa, Nucl. Phys. {\bf 37}, 411 (1962).
\bibitem{sm68} F.\,Smend, W.\,Weirauch, and W.-D.\,Schmidt-Ott, Z. Phys. {\bf 214}, 437 (1968).
\bibitem{NPA708} D.\,R.\,Tilley, C.M.\,Cheves, J.L.\,Godwina, G.M.\,Hale,
H.M.\,Hofmann, J.H.\,Kelley, C.G.\,Sheu, H.R.\,Weller, Nucl. Phys. {\bf A708}, 3 (2002).
\end{thebibliography}
\end{document}